\documentclass[aps,prl,reprint,groupedaddress,showpacs,longbibliography]{revtex4-1}

\usepackage{amssymb,amsmath}
\usepackage{graphicx}
\usepackage{xcolor}

\usepackage{siunitx}

\begin{document}

\title{Sound propagation in a uniform superfluid two-dimensional Bose gas}

\author{J.L. Ville}
\author{R. Saint-Jalm}
\author{\'E. Le Cerf}
\author{M. Aidelsburger$^{\dagger}$}
\author{S. Nascimb\`ene}
\author{J. Dalibard}
\author{J. Beugnon}

\email[]{beugnon@lkb.ens.fr}

\altaffiliation[$^\dagger$Present address: ]{Fakult\"at f\"ur Physik, Ludwig-Maximilians-Universit\"at M\"unchen, Schellingstr. 4, 80799 Munich, Germany.}

\affiliation{$^1$Laboratoire Kastler Brossel,  Coll\`ege de France, CNRS, ENS-PSL University, Sorbonne Universit\'e, 11 Place Marcelin Berthelot, 75005 Paris, France}

\date{\today}
\begin{abstract}
\noindent
In superfluid systems several sound modes can be excited, as for example first and second sound in liquid helium. Here, we excite propagating and standing waves in a uniform two-dimensional  Bose gas and we characterize the propagation of sound in both the superfluid and normal regime. In the superfluid phase, the measured speed of sound is well described by a two-fluid hydrodynamic model, and the weak damping rate is well explained by the scattering with thermal excitations. In the normal phase the sound becomes strongly damped due to a departure from hydrodynamic behavior.

\end{abstract}

\maketitle

Propagation of sound waves is at the heart of our understanding of quantum fluids. In liquid helium, the celebrated two-fluid model was confirmed by the observation of first and second sound modes \cite{Donnelly09}. There, first sound stands for the usual sound appellation, namely a density wave for which normal and superfluid fractions oscillate in phase. Second sound corresponds to  a pure entropy wave with no perturbation in density (normal and superfluid components oscillating out of phase), and is generally considered as a smoking gun of superfluidity. 

Sound wave propagation is also central to the study of dilute quantum gases, providing information on thermodynamic properties, relaxation mechanisms and superfluid behavior. In ultracold strongly interacting Fermi gases, the existence of first and second sound modes in the superfluid phase was predicted \cite{Taylor09} and observed in experiments \cite{Joseph07,Sidorenkov13}, with a behavior similar to liquid helium. In weakly interacting Bose-Einstein condensates (BECs), one still expects two branches of sound with speeds $c^{(1)}>c^{(2)}$ but the nature of first and second sound is strongly modified because of their large compressibility \cite{Griffin97}. While at zero temperature density perturbations propagate as Bogoliubov sound waves, at finite temperature we expect them to couple mostly to second sound -- a behavior contrasting with the case of liquid helium -- with a sound speed proportional to the square root of the superfluid fraction \cite{Griffin97,Ota18}.  Sound waves in an elongated  three-dimensional (3D) BEC were observed in Refs. \cite{Andrews97,Meppelink09} in a regime where the sound speed remains close to the Bogoliubov sound speed.

Propagation of sound in weakly interacting two-dimensional (2D) Bose gases was recently discussed in Ref.\,\cite{Ozawa14} using a hydrodynamic two-fluid model, predicting  the existence of first and second sound modes of associated speeds $c^{(1)}_{\rm HD}$ and $c^{(2)}_{\rm HD}$, respectively. In 2D Bose gases, superfluidity occurs via the Berezinskii-Kosterlitz-Thouless (BKT) mechanism \cite{Hadzibabic11}. The superfluid to normal transition is associated with a jump of the superfluid density that cannot be revealed from the thermodynamic properties of the gas. As the second sound speed is related to the superfluid fraction, one expects $c^{(2)}_{\rm HD}$ to remain non-zero just below the critical point of the superfluid to normal transition and to disappear just above the transition.

\begin{figure}[t!!]
\includegraphics[angle=0,width=8.6cm]{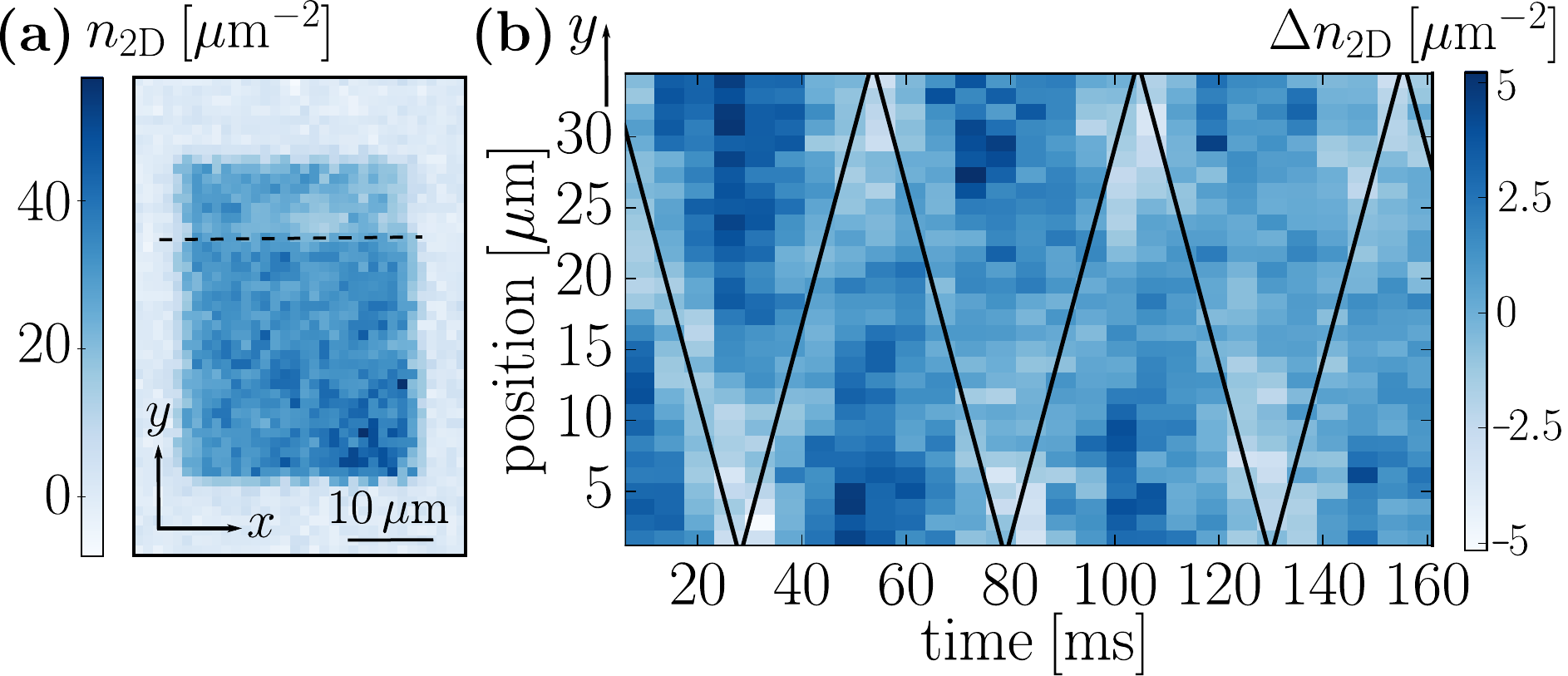}
\caption{Experimental protocol and observation of propagating waves. (a) Absorption image of the cloud perturbed by a local additional potential. The excitation is delimited by the horizontal dashed line and depletes the atomic density by a factor around $1/3$. (b) Example of time evolution of the variation of the density profile $n_{\rm 2D}$ with respect to its spatial mean value (integrated along $x$) obtained after abruptly removing the additional potential. For this example $T/T_{\rm c}=0.37(12)$ and $n_{\rm 2D}=29(3)\,\,\si{\micro m}^{-2}$. The position of the dip is fitted by a triangle function (black solid line) which gives, $c=$1.49(3) mm/s.}\label{fig1}
\end{figure}

In this Letter, we study the propagation of sound in a 2D uniform Bose gas. We observe a single density sound mode both in the superfluid and normal phases. Deep in the superfluid regime, the measured sound speed agrees well with the Bogoliubov prediction. We measure a weak damping rate compatible with Landau damping, a fundamental mechanism for the understanding of collective modes of superfluids at finite temperature \cite{Pitaevskii97}. For higher temperatures, we observe a decrease of the sound velocity consistent with the second sound speed variation predicted in Ref.\,\cite{Ozawa14} from two-fluid hydrodynamics. The damping of sound increases with temperature, and the sound propagation becomes marginal for temperatures close to the superfluid to normal transition. Above the critical point, we still observe strongly damped density waves, with no discernable discontinuity at the critical point. The discrepancy with the two-fluid model predictions could be due to a departure from hydrodynamic behavior, that manifests in our experiments as a strong damping of sound  around the critical point.

Our experimental setup has been described in Refs \cite{Ville17,Aidelsburger17}. Briefly, we confine $^{87}$Rb atoms in the $|F=1,m=0\rangle$ ground state into a 2D rectangular box potential of size $L_x \times L_y= 30(1) \times 38(1)\,\,\si{\micro m}$ (see Fig.\,\ref{fig1}a). The trapping potential is made by a combination of far-detuned repulsive optical dipole traps. The confinement along the vertical $z$ direction can be approximated by a harmonic potential of frequency $\omega_z/(2\pi)= 4.59(4)\,$kHz. We always operate in the quasi-2D regime where interaction and thermal energies are smaller than $\hbar \omega_z$. Collisions in our weakly-interacting Bose gas are characterized  by the effective coupling constant $g=\hbar^2 \tilde g /m =(\hbar^2/m) \sqrt{8\pi}\, a_{\rm s}/\ell_z$, where $a_s$ is the s-wave scattering length, $\ell_z=\sqrt{\hbar/(m\omega_z)}$ and $m$ the atomic mass \cite{Hadzibabic11}. With our confinement we have $\tilde g=0.16(1)$ \footnote{The value of $\tilde g$ is slightly modified by the effect of interactions. We estimate that $\tilde g$ varies by about 10\% for the range of surface densities explored in this work.}.
We control the temperature $T$ thanks to evaporative cooling by varying the height of the potential barrier providing the in-plane confinement. The surface density $n_{\rm 2D}$ of the cloud is varied from 10 to 80 $\si{\micro m}^{-2}$ by removing a controlled fraction of the atoms from our densest configuration \footnote{This removal is realized by a partial transfer of the atoms to the \unexpanded{$|F=2,m=0\rangle$} state with a microwave resonant field and a subsequent blasting of the transferred fraction with a resonant laser beam}. In the quasi-2D regime and for a given $\tilde g$, the equilibrium state of the cloud is only characterized by a dimensionless combination of $T$ and $n_{\rm 2D}$, thanks to an approximate scale-invariance \cite{Hadzibabic11}. In the following we use the ratio $T/T_{\rm c}$, where $T_{\rm c}=2\pi n_{\rm 2D} \hbar^2/[mk_{\rm B} \ln(380/ \tilde g)]$ is the calculated critical temperature for the BKT phase transition \cite{Prokofiev01}. We determine the ratio $T/T_{\rm c}$ by a method inspired from Ref.\,\cite{Hueck18} and based on a measurement of the equation of state of the system (see \cite{REFSM} for more details). In this work, we study Bose gases from the highly degenerate regime ($T/T_{\rm c} \approx 0.2$) to the normal regime ($T/T_{\rm c} \approx 1.4$).

We first investigate propagating waves which we excite by a density perturbation. Prior to evaporative cooling in the box potential, we apply to the cloud a repulsive potential, which creates a density dip on one side of the rectangle (see Fig.\,\ref{fig1}a). The extension of this dip is about 1/4 of the length of the box and its amplitude is chosen so that the density in this region is decreased by a factor of 1/3.  After equilibration, we abruptly remove the additional potential and monitor the propagation of this density dip. We show in Fig.\,\ref{fig1}b a typical time evolution of the density profile integrated along the transverse direction to the perturbation for a strongly degenerate gas \footnote{We did not observe any excitation along the $x$ direction.}.  In this regime, the density perturbation propagates at constant speed and bounces several times off the walls of the box. Using the calibrated size of the box, we extract a speed $c= 1.49(3)\,$mm/s. This value is slightly lower than the Bogoliubov sound speed $c_{\rm B}=\sqrt{gn_{\rm 2D}/m }=1.6(1)\,$mm/s expected at zero temperature for the measured density $n_{\rm 2D}=29(3)\,\,\si{\micro m}^{-2}$.  The measured speed is also close to the second sound mode velocity $c^{(2)}_{\rm HD}=1.4(1)\,$mm/s, estimated from two-fluid hydrodynamics at our experimental value of $T/T_{\rm c}=0.37(12)$ \cite{Ozawa14}. The first sound, expected to propagate at a much higher speed $c^{(1)}_{\rm HD}=3.3(3)\,$mm/s \cite{Ozawa14}, does not appear in our measurements that  feature a single wavefront only. The absence of first sound in our experiments can be explained by its very small coupling to density excitations  in a weakly interacting gas \cite{Ozawa14}.

\begin{figure}[t!!]
\includegraphics[width=8.6cm]{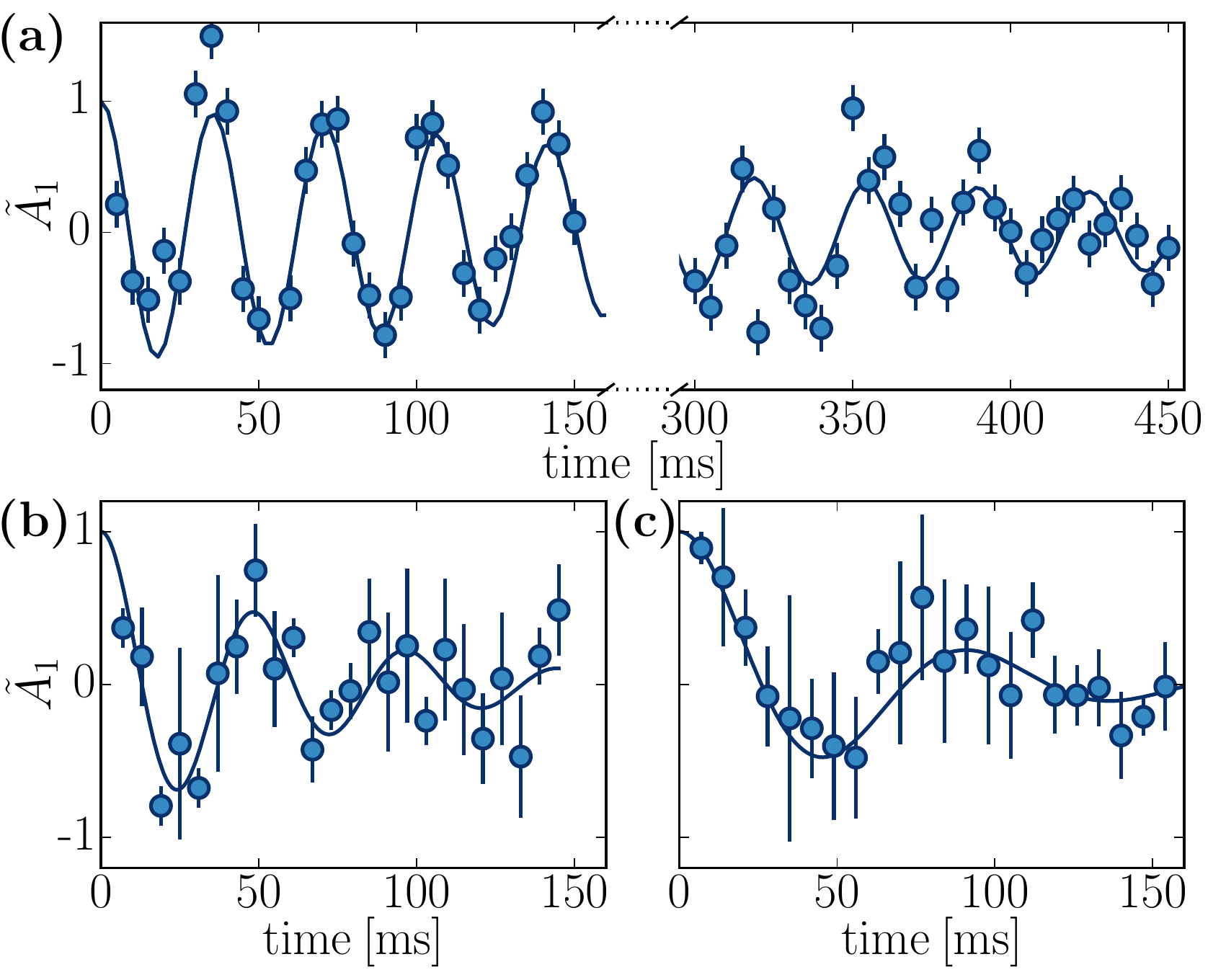}
\caption{Time evolution of the normalized amplitude of the lowest-energy mode for (a) $T/T_{\rm c}= 0.37(12)$, (b) $T/T_{\rm c}= 0.95(5)$, (c) $T/T_{\rm c}= 1.38(18)$. The solid line is a fit of an exponentially damped sinusoidal oscillation. For (b) and (c) graphs, each data point is the average of three measurements and the error bars represent the associated standard deviation. In (a) each point corresponds to a single measurement.}\label{fig2}
\end{figure}

In order to probe the role of the cloud degeneracy on the sound wave propagation, we vary both $n_{\rm 2D}$ and $T$. For each configuration, we excite the cloud with the protocol described above, while adjusting the intensity of the depleting laser beam to keep the density dip around 1/3 of non-perturbed density.  At lower degeneracies, sound waves are strongly damped and the aforementioned measurements of the density dip position become inadequate. We thus focus on the time evolution of the lowest-energy mode \footnote{A related experimental study of the evolution of the fundamental mode of a 3D uniform weakly interacting Bose gas can be found in Ref. \cite{Navon16}.}. We decompose the density profiles integrated along $x$ as
\begin{eqnarray}
n(y,t)=\bar n+ \sum_{j=1}^\infty A_j (t)\: \cos(j \pi y/L_y),
\end{eqnarray}
where $\bar n$ is the average density along $y$ and the $A_j$ are the amplitudes of the modes. The choice of the cosine basis ensures the cancellation of the velocity field on the edges of the box. Our excitation protocol mainly couples to the lowest energy modes. We keep the excitation to a low value to be in the linear regime while still observing a clear signal for the lowest-energy mode, which in return provides a too weak signal for a quantitative analysis of higher modes \footnote{The study of the second spatial mode gives oscillation frequencies that are in good approximation twice larger than the lowest-energy mode and thus results in very similar speeds of sounds. However, the damping rate of this mode is also larger (see Fig.\,\ref{fig4}) and we cannot robustly estimate its lifetime for our deliberately weak excitation protocol.}. For each duration of the evolution, we compute the overlap of the atomic density profile with the lowest-energy mode. Examples of the time evolution of the normalized amplitude $\tilde A_1(t)=A_1(t)/A_1(0)$ for different degrees of degeneracy are shown in Fig.\,\ref{fig2}. We observe damped oscillations with a damping rate increasing with $T/T_{\rm c}$. We fit the experimental data by an exponentially damped sinusoidal curve $e^{-\Gamma t/2}[\Gamma /2 \omega \sin(\omega t) +\cos(\omega t)] $ to determine the energy damping rate $\Gamma$ and the frequency $\omega$ \footnote{The choice of this oscillating function ensures a null derivative of the amplitude of the mode at $t=0$, when the potential creating the density dip is removed. This behavior is expected from the continuity of the wavefunction and of its derivative describing the state of the gas at $t=0$.}. We then determine the speed of sound $c= L_y \omega/\pi$ and the quality factor of this mode $Q=2\omega/\Gamma.$

\begin{figure}[t!!]
\includegraphics[width=8.6cm]{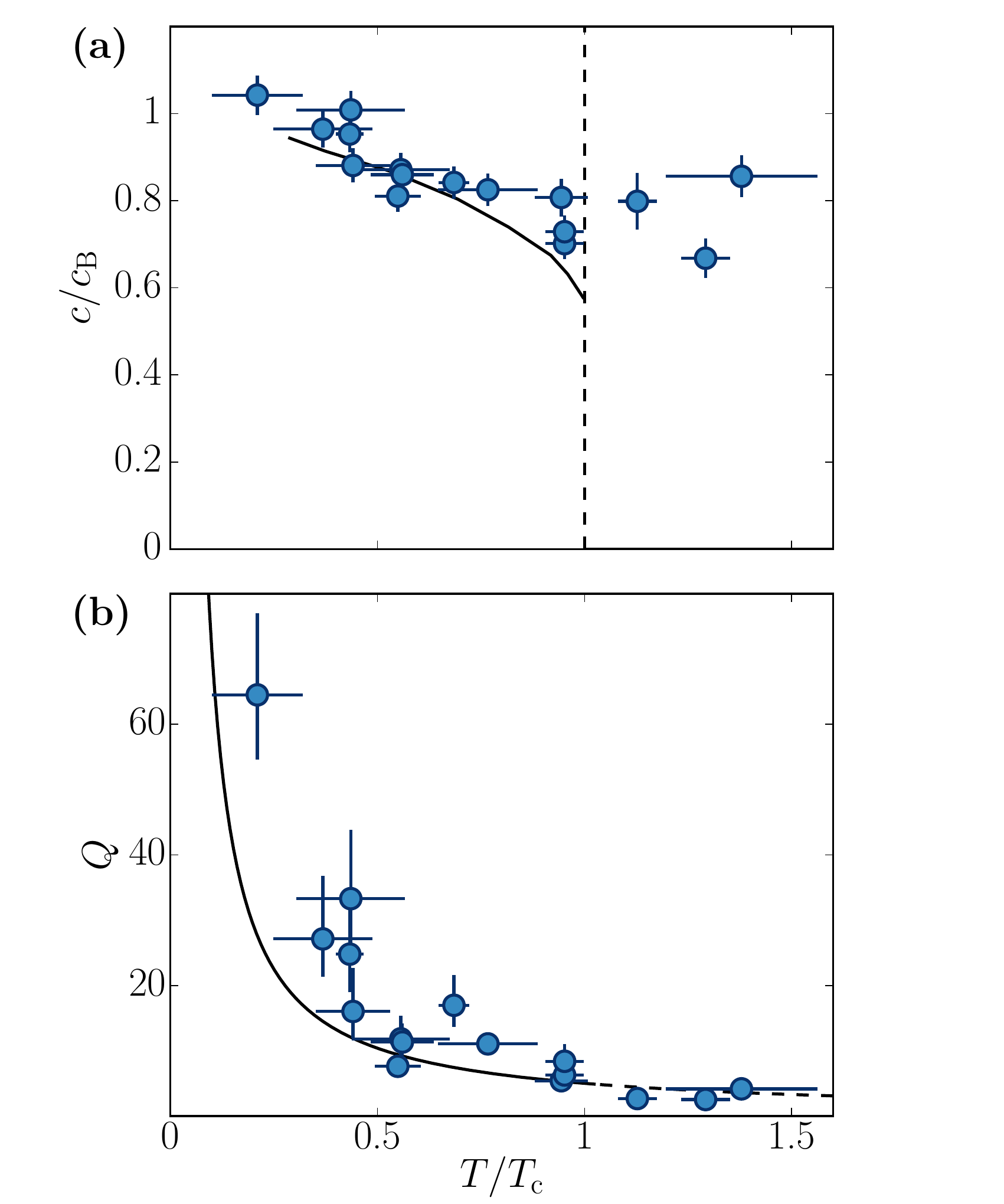}
\caption{Speed of sound and quality factor. (a) Measured speed of sound $c$ normalized to $c_{\rm B}$. The vertical dashed line shows the position of the critical point. The solid line shows the result from the two-fluid hydrodynamic model applied to the 2D Bose gas \cite{Ozawa14}. A fit to the data points below $T_{\rm c}$ by this hydrodynamic model with a free multiplicative factor shows that the measurements are globally 3\% above the theoretical prediction. This could correspond to a 6\% systematic error in the calibration of $n_{\rm 2D}$ used to determine $c_B \propto n_{\rm 2D}^{1/2}$. Our estimated uncertainty on $n_{\rm 2D}$ is on the order of $11\%$ (see Ref.\,\cite{REFSM}) and our measurements are thus compatible with the predicted value of the speed of second sound $c^{(2)}_{\rm HD}$. (b) Quality factor $Q=2\omega/\Gamma$ of the lowest-energy mode. The solid line is the prediction for Landau damping \cite{Chung09} (continued as a dashed line for $T>T_c$). For both graphs, the error bars represent the statistical uncertainty extracted from the fitting procedures used to determine $c$, $\Gamma$ and $T/T_{\rm c}$.}\label{fig3}
\end{figure}

We consolidate all our measurements of speed of sound and quality factors in Fig.\,\ref{fig3}. To facilitate comparison with theory, we show in Fig.\,\ref{fig3}a the values of $c$ normalized to $c_{\rm B}$. The non-normalized results are reported in Ref.\,\cite{REFSM} for completeness. In the temperature range $T \lesssim 0.9\,T_{\rm c}$, we measure weakly damped density oscillations, corresponding to a well-defined sound mode ($Q\gtrsim 10$). In this regime, we observe a significant decrease by about $\approx 25 \%$ of the sound velocity for increasing values of $T/T_{\rm c}$ . The measured velocities agree well with the prediction from  two-fluid hydrodynamics \cite{Ozawa14} combined with the equation of state of the 2D Bose gas \cite{Prokofiev02}. According to the analysis of \cite{Ozawa14} for weakly interacting gases, this variation is mainly due to the variation of the superfluid fraction $f_{\rm s}$ from $\approx 1$ at $T=0$ to $\approx 0.5$ close to $T=T_{\rm c}$ with the approximate scaling  $c^{(2)}_{\rm HD}\propto f_{\rm s}^{1/2}$ \cite{Ota18}. 

The measured quality factors (see Fig.\,\ref{fig3}b) compare rather well with the predictions of Ref.\,\cite{Chung09}, which calculates the decay of Bogoliubov quasi-particles via the Landau damping mechanism for a 2D uniform system \footnote{Note that Beliaev damping, another  mechanism for the decay of low-lying excitations, is absent for the first spatial mode of the box. Indeed, it corresponds to a decay of a low-lying excitation into two excitations with lower energies and thus does not exist for the lowest energy mode.}. Landau damping describes the  decay of low-lying collective excitations via scattering on  thermal excitations \cite{Pitaevskii97,Meppelink09b}. It predicts an increase of the quality factor when decreasing temperature due to the reduction of the number of thermal excitations available for scattering with the sound mode. While the agreement between our measurement and the Landau damping theory is fairly good for $T>0.5\,T_c$, we measure significantly larger quality factors for lower temperatures. This could be attributed to the collisionless nature of Landau prediction which does not apply well to our experimental situation. 

For temperatures above $0.9\,T_c$, we still observe sound waves, albeit with higher damping. In this regime, the measured sound speeds no longer match the two-fluid hydrodynamic model. The latter predicts a finite sound speed $c^{(2)}_{\rm HD}\simeq0.6\,c_{\rm B}$ for temperatures slightly below  the critical temperature, a value significantly below our measured values $c\simeq0.7-0.8\,c_{\rm B}$. More strikingly, it predicts a disappearance of second sound for $T>T_c$, while we still observe sound waves with a velocity comparable to $c^{(2)}_{\rm HD}$. The discrepancy between the hydrodynamic prediction and our measurements can be explained by the strong damping of the measured density oscillations. Indeed, for quality factors of order 1, dissipation effects are no longer perturbative, making  dissipationless hydrodynamic models less relevant. In the normal phase, we expect hydrodynamics  to describe well the propagation of a sound wave when the collision rate $\Gamma_{\mathrm{coll}}$ between particles largely exceeds the oscillation frequency $\omega$. Assuming quasi-2D kinematics, we estimate $\Gamma_{\mathrm{coll}}/\omega$ to be in the range $1.6-3.4$ \footnote{For a thermal gas evolving in a quasi-2D geometry, the collision rate reads $\Gamma_{\mathrm{coll}}=\hbar\tilde g^2 n/(2m)$ \cite{Petrov01}. We did not include the bosonic enhancement factor 2 as it should not be relevant for our temperature range, where we expect reduced density fluctuations even in the normal phase \cite{Prokofiev02}.}. We conclude that, in the normal phase, the gas dynamics is not expected to follow the hydrodynamics prediction, which could explain our observations \footnote{The existence of a sound mode in an interacting but collisionless cloud is still expected (S. Stringari, private communication). The role of interactions remains indeed important in the normal phase, even if hydrodynamics does not apply, because the healing length of the cloud is much smaller than the size of the box potential.}.

\begin{figure}[t!!]
\includegraphics[width=8.6cm]{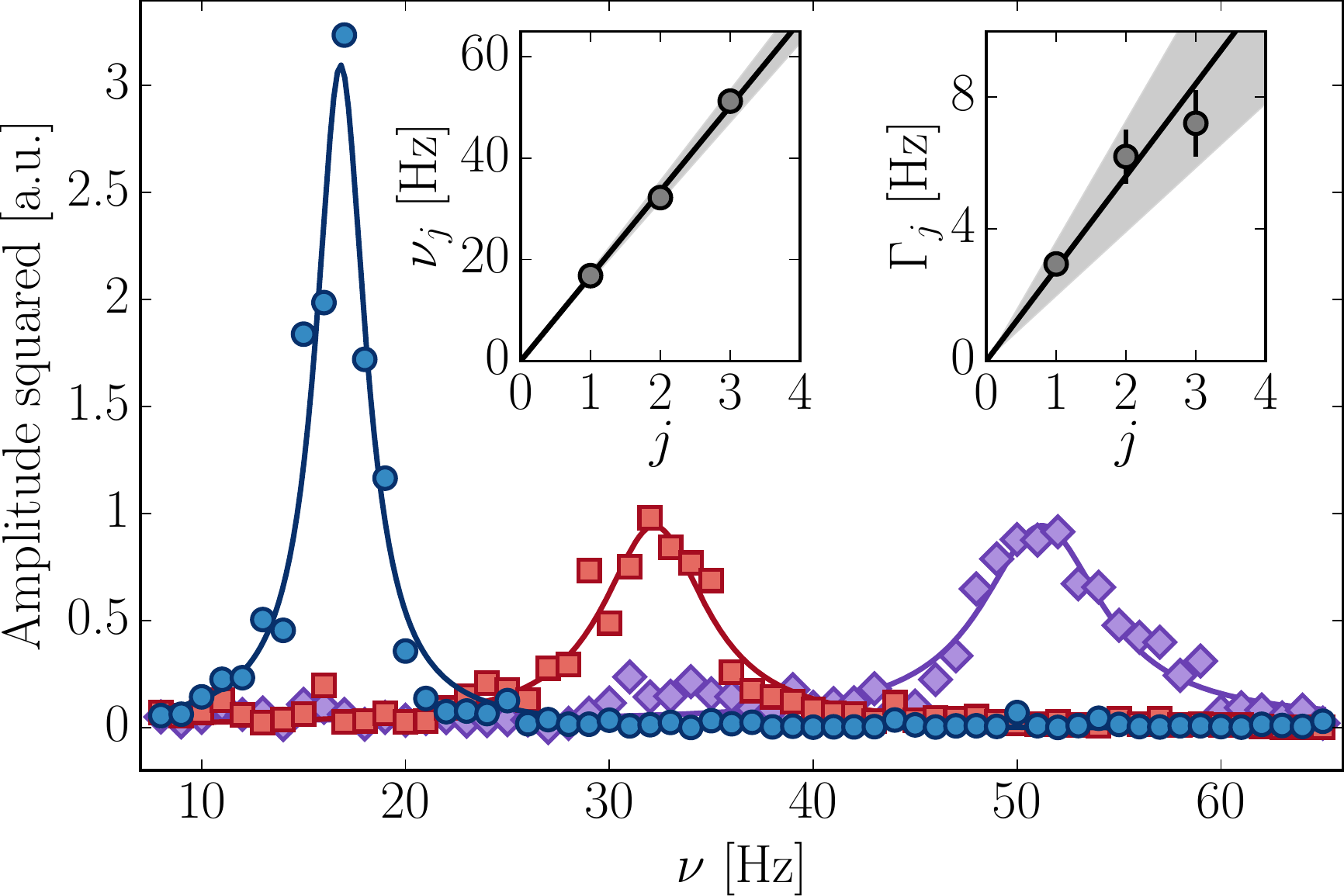}
\caption{Observation of standing waves in the box potential. Contribution of the three lowest-energy modes to the amplitude of the density modulation: $j=1$ (circles), $j=2$ (squares), $j=3$ (diamonds).  The solid lines are Lorentzian fits. The two insets show the resonance frequencies $\nu_j$ and the full widths at half maximum $\Gamma_j$ resulting from these fits. The solid lines in the insets are linear fit to the data and the shaded areas represent the uncertainty on the fitted slope. From the slope $c/(2L_y)$ of the fit to the resonance frequencies, we find $c=1.90(9)\,$mm/s. For this specific experiment, the length of the cloud is $L_y=57(1)\,\si{\micro m}$ and the degree of degeneracy is $T/T_{\rm c}=0.41(7)$.}\label{fig4}
\end{figure}

In the highly degenerate regime, the low damping rate allows us to observe standing waves. To study them, we modulate sinusoidally the amplitude of the potential creating the dip of density on one edge of the box \footnote{A similar protocol has been used in \cite{Wang15} to excite a degenerate Bose gas in a ring geometry.}. After $\approx$1\,s we extract, for each frequency $\nu$ of the excitation, the amplitude of the (time-dependent) density modulation induced on the cloud (see Ref.\,\cite{REFSM} for details). We show  in Fig.\,\ref{fig4} the contribution of the three lowest-energy modes to the amplitude of the modulation as a function of the excitation frequency. For each mode $j$ we observe a clear resonance peak centered at a frequency $\nu_j$. We display in the insets the resonance frequencies and width of the modes. The $\nu_j$'s are equally spaced, as confirmed by the linear fit. In addition, the right inset shows the widths of the peaks. They also increase approximately linearly with $j$ \footnote{Because of the finite duration of the excitation (1\,s), the width of the peaks is Fourier limited at a typical width of 1\,Hz, which should be taken into account for a more quantitative analysis.}, meaning that the quality factor associated to these peaks is almost the same, as expected for Landau damping.

In conclusion, we have reported the first measurement of second sound velocity and the associated damping in a uniform 2D quantum fluid, and we have characterized their variation with temperature. Surprisingly, this sound mode extends to above the critical temperature and may corresponds to a collisionless mode. This work focuses on a weakly interacting Bose gas which features a large compressibility compared to liquid helium or strongly interacting Fermi gases. A natural extension of this work would thus be to investigate second sound propagation for increasing interactions \cite{Ota18}. It would also be interesting to investigate first sound, e.g. by applying a localized temperature excitation \cite{Sidorenkov13}. During the completion of this work we were informed that a related study with a homogeneous 3D Fermi gas was currently performed at MIT \footnote{M. Zwierlein, talk at the BEC 2017 Frontiers in Quantum Gases, Sant Feliu de Guixols}.

\begin{acknowledgments}
$^\dagger$Present address: Fakult\"at f\"ur Physik, Ludwig-Maximilians-Universit\"at M\"unchen, Schellingstr. 4, 80799 Munich, Germany. This work is supported by DIM NanoK and ERC (Synergy UQUAM). This project has received funding from the European Union's Horizon 2020 research and innovation programme under the Marie Sk\l{}odowska-Curie grant agreement N$^\circ$ 703926. We thank S. Stringari, L. Pitaevskii, M. Ota, N. Proukakis, F. Dalfovo, F. Larcher and P.C.M. Castilho for fruitful discussions, M. Villiers for experimental assistance and F. Gerbier, R. Lopes and M. Zwierlein for their reading of the manuscript.

J.L.V. and R.S.J. contributed equally to this work.
\end{acknowledgments}

\bibliography{soundbib}

\end{document}